\documentclass{JHEP3}

\title{Study of the $\eta$-$\eta^\prime$ system in the two mixing angle scheme}
\author{
Rafel Escribano\\
Grup de F\'{\i}sica Te\`orica and IFAE,
Universitat Aut\`onoma de Barcelona,
E-08193 Bellaterra (Barcelona), Spain\\
E-mail: \email{Rafel.Escribano@ifae.es}
}
\author{
Jean-Marie Fr\`ere\\
Service de Physique Th\'eorique,
CP225 Universit\'e Libre de Bruxelles,
Bld.~du Triomphe, 1050 Brussels, Belgium\\
E-mail: \email{frere@ulb.ac.be}
}
\abstract{
An analysis of various decay processes is performed using
the two mixing angle description of the $\eta$-$\eta^\prime$ system,
incorporating the link to the gluonic sector through anomalies.
The agreement is excellent.
For comparison with previous works, our results are expressed both in the
``octet-singlet''  and in the ``quark-flavour'' basis.
It turns out that at the present experimental accuracy, 
the two angles are significantly different in the former, but not in the latter basis. 
The implications of our analysis for the Large $N_c$ $\chi$PT predictions are also discussed.
}
\preprint{UAB--FT--576\\
          ULB-TH/04-32}

\begin{document}
\section{Introduction}
\label{intro}
The phenomenon of mixing in the $\eta$-$\eta^\prime$ system has attracted much attention
since the advent of $SU_f(3)$ symmetry.
Conventionally, one regards $\eta$ and $\eta^\prime$ as linear combinations of octet and singlet
basis states, parametrized by a mixing angle $\theta_P$.
A determination of its value can be achieved either diagonalizing a mass matrix,
as done in Chiral Perturbation Theory ($\chi$PT), or from phenomenology \cite{Eidelman:2004wy}.
Several exhaustive analyses surveying many different processes have been performed along the years
with values for the mixing angle ranging from $-24^\circ$ to $-10^\circ$
\cite{Gilman:1987ax}--\cite{Herrera-Siklody:1997kd}.
In all these previous analyses the dependence with energy of the $\eta$-$\eta^\prime$ mixing angle
has been neglected.
A recent analysis assuming such an energy dependence is also present in the literature
\cite{Escribano:1999nh}.

The $\eta$-$\eta^\prime$ mixing is strongly connected with the $U(1)$ anomaly of QCD.
Recently, it has pointed out that an extension of $\chi$PT taking into account the effects
of the chiral anomaly through a perturbative expansion leads to a description of the
$\eta$-$\eta^\prime$ system in terms of two mixing angles \cite{Leutwyler:1997yr}.
In this framework, the four pseudoscalar decay constants associated to the matrix elements of
axial-vector currents are written in terms of two basic decay constants $f_8, f_0$
and two angles $\theta_8, \theta_0$ \cite{Kaiser:1998ds}.
Alternative approaches also including the chiral anomaly \textit{ab initio}
but using a more conventional one mixing angle scheme are discussed in
Refs.~\cite{Ball:1995zv,Akhoury:1987ed,Kiselev:1992ms,DeFazio:2000my,Gerard:2004gx}.
On the phenomenological side, the idea of using a two mixing angle scheme has been explored 
in Refs.~\cite{Feldmann:1998vh,Benayoun:1999au}.

The aim of this work is to perform an updated phenomenological analysis
of various decay processes using the two mixing angle description of the $\eta$-$\eta^\prime$ system.
The analysis is carried out in the octet-singlet and quark-flavour basis
and our theoretical predictions compared with the latest experimental data.
The comparison will serve us to check the validity of the two mixing angle scheme and its improvement
over the standard one angle picture.
The analysis also tests the sensitivity to the mixing angle schemes.

In Section \ref{notation}, we shortly introduce the notation used in the analysis.
In Section \ref{expvaluesangles},
we compute the radiative decays $(\eta, \eta^\prime)\to\gamma\gamma$
and the ratio $R_{J/\psi}\equiv\Gamma(J/\psi\to\eta^\prime\gamma/\eta\gamma)$
in the two mixing angle scheme of both the octet-singlet and quark-flavour basis
in order to obtain in each basis the preferred values for the mixing angles involved in the analysis.
Section \ref{VPgamma} is devoted to the consequences of our approach in the context of the
radiative decays of the lowest-lying vector and pseudoscalar mesons,
$V\to P\gamma$ and $P\to V\gamma$, respectively.
In Section \ref{discussion}, we compare our best results for the mixing parameters
with the theoretical expectations of Large $N_c$ $\chi$PT and previous phenomenological analyses.
The implications of our analysis on these Large $N_c$ $\chi$PT predictions are also discussed.
Finally, in Section \ref{conclusions}, we present our conclusions.

\section{Notation}
\label{notation}
The decay constants of the $\eta$-$\eta^\prime$ system in the octet-singlet basis
$f_P^a\ (a=8,0; P=\eta,\eta^\prime)$ are defined as\footnote{
The axial-vector currents are defined as
$A_\mu^a=\bar q\gamma_{\mu}\gamma_{5}\frac{\lambda^a}{\sqrt{2}}q$
with the normalization convention $f_{\pi}=\sqrt{2}F_{\pi}=130.7$ MeV.}
\begin{equation}
\label{decaycon}
\langle 0|A_\mu^a|P(p)\rangle=i f_P^a p_\mu\ ,
\end{equation}
where $A_\mu^{8,0}$ are the octet and singlet axial-vector currents whose
divergences are
\begin{equation}
\label{divaxialos}
\begin{array}{l}
\partial^\mu A_\mu^8=\frac{2}{\sqrt{6}}
(m_u\bar u i\gamma_5 u+m_d\bar d i\gamma_5 d-
2m_s\bar s i\gamma_5 s)\ ,\\[2ex]
\partial^\mu A_\mu^0=\frac{2}{\sqrt{3}}
(m_u\bar u i\gamma_5 u+m_d\bar d i\gamma_5 d+m_s\bar s i\gamma_5 s)+
\frac{1}{\sqrt{3}}\frac{3\alpha_s}{4\pi} G_{\mu\nu}^a\tilde G^{a,\mu\nu}\ ,
\end{array}
\end{equation}
where $G^a_{\mu\nu}$ is the gluonic field-strength tensor and
$\tilde G^{a,\mu\nu}\equiv\frac{1}{2}\epsilon^{\mu\nu\alpha\beta}G^a_{\alpha\beta}$
its dual.
The divergence of the matrix elements (\ref{decaycon}) are then written as
\begin{equation}
\label{divdecaycon}
\langle 0|\partial^\mu A_\mu^a|P\rangle=f_P^a m_P^2\ ,
\end{equation}
where $m_P$ is the mass of the pseudoscalar meson.

Each of the two mesons $P=\eta, \eta^\prime$ has both octet and singlet components,
$a=8, 0$.
Consequently, Eq.~(\ref{decaycon}) defines \emph{four independent} decay constants.
Following the convention of Refs.~\cite{Leutwyler:1997yr,Kaiser:1998ds}
the decay constants are parameterized in terms of two basic decay constants
$f_{8}, f_{0}$ and two angles $\theta_{8}, \theta_{0}$
\begin{equation}
\label{defdecaybasisos}
\left(
\begin{array}{cc}
f^8_\eta & f^0_\eta \\[1ex]
f^8_{\eta^\prime} & f^0_{\eta^\prime}
\end{array}
\right)
=
\left(
\begin{array}{cc}
f_8 \cos\theta_8 & -f_0 \sin\theta_0 \\[1ex]
f_8 \sin\theta_8 &  f_0 \cos\theta_0
\end{array}
\right)\ .
\end{equation}
This parametrization is the most general involving two independent axial-vector currents
and two different physical states.

Neglecting the contribution of the {\it up} and {\it down} quark masses,
as in Ref.~\cite{Akhoury:1987ed}, the matrix elements of the chiral anomaly between
the vacuum and $(\eta, \eta^\prime)$ states are
\begin{equation}
\label{chiralanomalyos}
\begin{array}{l}
\langle 0|\frac{3\alpha_s}{4\pi} G\tilde G|\eta\rangle=
\sqrt{\frac{3}{2}}m_\eta^2 
(f_8 \cos\theta_8-\sqrt{2} f_0 \sin\theta_0)\ ,\\[1ex]
\langle 0|{\frac{3\alpha_s}{4\pi}} G\tilde G|\eta^\prime\rangle=
\sqrt{\frac{3}{2}}m_{\eta^\prime}^2 
(f_8 \sin\theta_8+\sqrt{2} f_0 \cos\theta_0)\ .
\end{array}
\end{equation}

Analogously, in the quark-flavour basis the decay constants are parameterized in terms of
$f_{q}, f_{s}$ and $\phi_{q}, \phi_{s}$:
\begin{equation}
\label{defdecaybasisqf}
\left(
\begin{array}{cc}
f^q_\eta & f^s_\eta \\[1ex]
f^q_{\eta^\prime} & f^s_{\eta^\prime}
\end{array}
\right)
=
\left(
\begin{array}{cc}
f_q \cos\phi_{q} & -f_s \sin\phi_{s}\\[1ex]
f_q \sin\phi_{q} &  f_s \cos\phi_{s}
\end{array}
\right)\ ,
\end{equation}
and the non-strange and strange axial-vector currents are defined as
\begin{equation}
\label{divaxialqfos}
    \begin{array}{l}
    A_\mu^q=\frac{1}{\sqrt{2}}(\bar u\gamma_{\mu}\gamma_{5}u+
                               \bar d\gamma_{\mu}\gamma_{5}d)
           =\frac{1}{\sqrt{3}}(A_\mu^8+\sqrt{2}A_\mu^0)\ ,\\[2ex]
    A_\mu^s=\bar s\gamma_{\mu}\gamma_{5}s
           =\frac{1}{\sqrt{3}}(A_\mu^0-\sqrt{2}A_\mu^8)\ . 
    \end{array}
\end{equation}
The divergences of these currents are
\begin{equation}
\label{divaxialqf}
\begin{array}{l}
\partial^\mu A_\mu^q=\sqrt{2}
(m_u\bar u i\gamma_5 u+m_d\bar d i\gamma_5 d)+
\frac{\sqrt{2}}{3}\frac{3\alpha_s}{4\pi} G_{\mu\nu}^a\tilde G^{a,\mu\nu}\ ,\\[2ex]
\partial^\mu A_\mu^s=2 m_s\bar s i\gamma_5 s+
\frac{1}{3}\frac{3\alpha_s}{4\pi} G_{\mu\nu}^a\tilde G^{a,\mu\nu}\ ,
\end{array}
\end{equation}
and therefore the matrix elements of the chiral anomaly in this basis are
\begin{equation}
\label{chiralanomalyqf}
\langle 0|\frac{3\alpha_s}{4\pi} G\tilde G|\eta\rangle=
\frac{3}{\sqrt{2}}m_\eta^2 f_q \cos\phi_{q}\ ,\quad
\langle 0|{\frac{3\alpha_s}{4\pi}} G\tilde G|\eta^\prime\rangle=
\frac{3}{\sqrt{2}}m_{\eta^\prime}^2 f_q \sin\phi_{q}\ .
\end{equation}

\section{Experimental values for the $\theta_{8,0}$ and $\phi_{q,s}$ mixing angles}
\label{expvaluesangles}
In order to reach some predictions from our two mixing angle analysis
we must first know the values of $\theta_8$ and $\theta_0$
preferred by the experimental data.
We will use as constraints\footnote{
We choose such constrains because those decays are well understood
in terms of the electromagnetic anomaly (see e.g.~Ref.~\protect\cite{Donoghue:dd}).} 
the experimental decay widths of 
$(\eta, \eta^\prime)\rightarrow\gamma\gamma$ \cite{Eidelman:2004wy}
\begin{equation}
\label{expwidths}
\begin{array}{c}
\Gamma(\eta\rightarrow\gamma\gamma)=(0.510\pm 0.026)\ \mbox{keV}\ ,\\[1ex]
\Gamma(\eta^\prime\rightarrow\gamma\gamma)=(4.29\pm 0.15)\ \mbox{keV}\ .
\end{array}
\end{equation}
Analogously to the $\pi^0\rightarrow\gamma\gamma$ case,
one assumes that the interpolating fields $\eta$ and $\eta^\prime$ can be
related with the axial-vector currents
(see e.g.~Refs.~\cite{Akhoury:1987ed,Kiselev:1992ms}) 
in the following way:
\begin{equation}
\label{interfields}
\begin{array}{c}
\eta(x)=\frac{1}{m_\eta^2}
\frac{f^0_{\eta^\prime}\partial^\mu A_\mu^8(x)-
      f^8_{\eta^\prime}\partial^\mu A_\mu^0(x)}
{f^0_{\eta^\prime}f^8_\eta-f^8_{\eta^\prime}f^0_\eta}\ ,\\[2ex]
\eta^\prime(x)=\frac{1}{m_{\eta^\prime}^2}
\frac{f^0_\eta\partial^\mu A_\mu^8(x)-
      f^8_\eta\partial^\mu A_\mu^0(x)}
{f^0_\eta f^8_{\eta^\prime}-f^8_\eta f^0_{\eta^\prime}}\ .
\end{array}
\end{equation}
This leads to\footnote{Note that if one assumes a one mixing angle scheme
$(\theta_8=\theta_0\equiv\theta)$ the standard result is obtained
(see e.g.~Ref.~\protect\cite{Donoghue:dd}).}
\begin{equation}
\label{theowidths}
\begin{array}{c}
\Gamma(\eta\rightarrow\gamma\gamma)=
\frac{\alpha^2 m_\eta^3}{96\pi^3}\left(
\frac{f^0_{\eta^\prime}-2\sqrt{2}f^8_{\eta^\prime}}
{f^0_{\eta^\prime}f^8_\eta-f^8_{\eta^\prime}f^0_\eta}\right)^2=
\frac{\alpha^2 m_\eta^3}{96\pi^3}\left(
\frac{c\theta_0/f_8-2\sqrt{2}s\theta_8/f_0}
{c\theta_0 c\theta_8+s\theta_8 s\theta_0}\right)^2\ ,\\[2ex]
\Gamma(\eta^\prime\rightarrow\gamma\gamma)=
\frac{\alpha^2 m_{\eta^\prime}^3}{96\pi^3}\left(
\frac{f^0_\eta-2\sqrt{2}f^8_\eta}
{f^0_\eta f^8_{\eta^\prime}-f^8_\eta f^0_{\eta^\prime}}\right)^2=
\frac{\alpha^2 m_{\eta^\prime}^3}{96\pi^3}\left(
\frac{s\theta_0/f_8+2\sqrt{2}c\theta_8/f_0}
{c\theta_0 c\theta_8+s\theta_8 s\theta_0}\right)^2\ .
\end{array}
\end{equation}
Because of the four unknown parameters 
($\theta_8, \theta_0, f_8$ and $f_0$) that appear in 
Eq.~(\ref{theowidths}), in order to get their allowed values we need two 
additional constraints (apart from the experimental constraints in 
Eq.~(\ref{expwidths})).
On the one hand, we will use the well established prediction of 
$\chi$PT: $f_8=1.28 f_\pi$ ($f_\pi=130.7$ MeV) as a theoretical constrain
(see variants on this later).
On the other hand, we will use the experimental value of the ratio
\cite{Eidelman:2004wy}
\begin{equation}
\label{RJpsiexp}
R_{J/\psi}\equiv
\frac{\Gamma(J/\psi\rightarrow\eta^\prime\gamma)}
     {\Gamma(J/\psi\rightarrow\eta\gamma)}
=5.0\pm 0.6\ .
\end{equation}
According to Ref.~\cite{Novikov:uy}, the radiative $J/\psi\rightarrow P\gamma$ decays
are dominated by non-perturbative gluonic matrix elements
(see Ref.~\cite{Ball:1995zv} for further comments on the accuracy of this statement):
\begin{equation}
\label{RJpsitheo}
R_{J/\psi}=\left|
\frac{\langle 0|G\tilde G|\eta^\prime\rangle}{\langle 0|G\tilde G|\eta\rangle}
\right|^2\left(\frac{p_{\eta^\prime}}{p_\eta}\right)^3\ ,
\end{equation}
where $p_P=M_{J/\psi}(1-m_P^2/M_{J/\psi}^2)/2$ is the three-momentum of the 
$P$-meson in the rest frame of the decaying $J/\psi$ (with mass $M_{J/\psi}$). 
Using Eq.~(\ref{chiralanomalyos}) one gets
\begin{equation}
\label{RJpsiourtheo}
\begin{array}{rcl}
R_{J/\psi}&=&
\left|
\frac{m_{\eta^\prime}^2(f^8_{\eta^\prime}+\sqrt{2}f^0_{\eta^\prime})}
     {m_\eta^2(f^8_\eta+\sqrt{2}f^0_\eta)}
\right|^2
\left(\frac{p_{\eta^\prime}}{p_\eta}\right)^3\\[2ex]
&=&
\left[\frac{m_{\eta^\prime}^2(f_8\sin\theta_8+\sqrt{2}f_0\cos\theta_0)}
           {m_\eta^2(f_8\cos\theta_8-\sqrt{2}f_0\sin\theta_0)}\right]^2
\left(\frac{p_{\eta^\prime}}{p_\eta}\right)^3\ .
\end{array}
\end{equation}
Comparing the experimental values of 
$\Gamma(\eta,\eta^\prime\rightarrow\gamma\gamma)$ and $R_{J/\psi}$ with the
theoretical predictions shown in Eqs.~(\ref{theowidths}) and 
(\ref{RJpsiourtheo}), one obtains
\begin{equation}
\label{fitresult128}
\theta_8=(-22.2\pm 1.8)^\circ\ ,\quad
\theta_0=(-8.7 \pm 2.1)^\circ\ ,\quad 
f_0=(1.18\pm 0.04) f_\pi\ .
\end{equation}
Using instead of $f_8=1.28 f_\pi$ the prediction of Large $N_{c}$ $\chi$PT
\cite{Kaiser:2000gs, Herrera-Siklody:1996pm}:
$f_8=1.34 f_\pi$ \cite{Kaiser:1998ds}, we would have found
\begin{equation}
\label{fitresult134}
\theta_8=(-22.9\pm 1.8)^\circ\ ,\quad
\theta_0=(-6.9 \pm 2.0)^\circ\ ,\quad 
f_0=(1.20\pm 0.04) f_\pi\ .
\end{equation}
The previous results are in agreement with the expectations of Large $N_{c}$ $\chi$PT
\cite{Leutwyler:1997yr,Kaiser:1998ds} and with phenomenological analyses
\cite{Feldmann:1998vh,Feldmann:1998sh,Benayoun:2003we} 
(see also Ref.~\cite{Feldmann:1999uf} for a comparison of various analyses).
Notice that the numerical value of $f_0$ cannot in principle be determined unambiguously,
because it depends on the renormalization scale \cite{Espriu:1982bw,Shore:1991dv}.
However, the scale dependence of the anomalous dimension of the singlet axial-vector
current is very weak \cite{Leutwyler:1997yr}.
Therefore, in our numerical analysis we neglect that dependence and consider $f_0$
as a constant.
The results of Eqs.~(\ref{fitresult128}) and (\ref{fitresult134})
constitute the first result of the present analysis.
It is worth noting that the $\theta_8$ and $\theta_0$ 
mixing angle values are different at the $3\sigma$ level, while the values of
the pseudoscalar decay constant $f_0$ remain compatible with other results.

In order to check the consistency of these results one can perform the same analysis
but fixing $\theta_{8}=\theta_{0}\equiv\theta$, \textit{i.e.~}a one mixing angle scheme.
Keeping $f_8=1.28 f_\pi$ and using $\eta,\eta^\prime\rightarrow\gamma\gamma$
in order to fit $f_{0}$ and $\theta$, one gets:
$\theta=(-22.2\pm 1.7)^\circ$ and $f_{0}=(1.07\pm 0.04) f_\pi$
---or $\theta=(-22.9\pm 1.7)^\circ$ and $f_{0}=(1.06\pm 0.04) f_\pi$ if $f_8=1.34 f_\pi$.
In these cases, however, the predicted value for $R_{J/\psi}\simeq 2.1$ (or $\simeq 1.7$)
is incompatible with the experimental result in Eq.~(\ref{RJpsiexp}).
Alternatively, leaving $f_{8}$ as a free parameter one gets:
$\theta=(-17.7\pm 1.2)^\circ, f_{8}=(1.02\pm 0.06) f_\pi$ and $f_{0}=(1.09\pm 0.04) f_\pi$.
In this other case, $R_{J/\psi}\simeq 3.5$ and the value of $f_{8}$ is incompatible
at the $3\sigma$ level with the predictions of $\chi$PT and Large $N_{c}$ $\chi$PT.

Therefore, a two mixing angle scheme leads to significantly different mixing angles in the
octet-singlet basis to describe the radiative $\eta,\eta^\prime\rightarrow\gamma\gamma$ decays
and the ratio $R_{J/\psi}$ simultaneously.

In the quark-flavour basis (see the Appendix for the corresponding theoretical expressions),
a simultaneous fit of
$\eta,\eta^\prime\rightarrow\gamma\gamma$ and $R_{J/\psi}$ with $f_{q}=f_{\pi}$
suggests that, at the current experimental precision, 
$\phi_{q}\simeq 39.8^\circ$ and $\phi_{s}\simeq 38.6^\circ$.
Therefore, in this particular basis only one mixing angle seems needed to describe
the data at the current experimental accuracy.
If, for the exercise, we force this equality (which is \emph{not} based in theory),
$\phi_{q}=\phi_{s}\equiv\phi$, the result of the fit is
\begin{equation}
\label{fitresultquark}
f_q=(1.07\pm 0.03) f_\pi\ ,\quad
f_s=(1.37\pm 0.27) f_\pi\ ,\quad
\phi=(39.0\pm 1.7)^\circ\ ,
\end{equation}
which is in agreement with earlier results \cite{Feldmann:1998vh}.
The value of the mixing angle $\phi$ is fixed from the ratio $R_{J/\psi}$.
Keeping however $f_q=f_\pi$, as predicted by the OZI-rule, leads to a poor fit with
$f_s=(1.39\pm 0.26) f_\pi$ and $\phi=(39.8\pm 1.5)^\circ$.

Let us mention at this stage that the fact of describing in terms of one or two angles
is in itself not a relevant theoretical consideration,
and that the resulting physical description should of course not depend upon the choice
of the initial basis.
In a general formulation, there are indeed two angles.
Furthermore, the choice of the starting point (octet-singlet or magic mixing) is in no way obvious.
The relevant symmetry, at least for the $\eta$-$\eta^\prime$ part of the present paper, is $U(3)$,
and subject to two types of breakings, one via the strong anomaly
(inducing the $SU(3)$-singlet term, which can also be seen as an OZI-rule violation),
the other via the strange quark current mass.
The effect of both breakings is similar in size in the mass matrix,
leading to a mixing structure somewhere between the two basis.
The $J/\psi$ decays are of course also clear OZI violations in this framework.

\section{$V$-$P$ electromagnetic form factors in the two mixing angle scheme}
\label{VPgamma}
In this Section, we want to extend our analysis to the $V$-$P$ electromagnetic form factors.
In particular, we are interested in the couplings of the radiative decays of
lowest-lying vector mesons, $V\rightarrow(\eta,\eta^\prime)\gamma$, and of the 
radiative decays $\eta^\prime\rightarrow V\gamma$, with $V=\rho, \omega, \phi$.
In order to predict such couplings we follow closely the method presented in
Ref.~\cite{Ball:1995zv} where the description of the light vector meson decays is based on
their relation with the $AVV$ triangle anomaly, $A$ 
and $V$ being an axial-vector and a vector current respectively.
The approach both includes $SU_f(3)$ breaking effects and fixes the vertex
couplings $g_{VP\gamma}$ as explained below.

In that framework, one starts considering the correlation function
\begin{equation}
\label{corrfun}
i\int d^4x e^{iq_1 x}
\langle P(q_1+q_2)|TJ_\mu^{\rm EM}(x)J_\nu^V(0)|0\rangle=
\epsilon_{\mu\nu\alpha\beta}q_1^\alpha q_2^\beta F_{VP\gamma}(q_1^2,q_2^2)\ ,
\end{equation}
where the currents are defined as
\begin{equation}
\label{defcurr}
\begin{array}{c}
J_\mu^{\rm EM}=\frac{2}{3}\bar u\gamma_\mu u-\frac{1}{3}\bar d\gamma_\mu d-
               \frac{1}{3}\bar s\gamma_\mu s\ ,\\[1ex]
J_\mu^{\rho,\omega}=
\frac{1}{\sqrt{2}}(\bar u\gamma_\mu u\mp\bar d\gamma_\mu d)
\quad\mbox{and}\quad J_\mu^\phi=-\bar s\gamma_\mu s\ .
\end{array}
\end{equation}
The form factors values $F_{VP\gamma}(0,0)$ are fixed by the $AVV$ triangle 
anomaly (one $V$ being an electromagnetic current), and are written in terms 
of the pseudoscalar decay constants and the $\phi$-$\omega$ mixing angle 
$\theta_V$ as\footnote{
The $\phi$-$\omega$ mixing angle in the octet-singlet basis is defined as
\[
\begin{array}{l}
    \phi=\cos\theta_V\omega_8-\sin\theta_V\omega_0\ ,\\[1ex]
    \omega=\sin\theta_V\omega_8+\cos\theta_V\omega_0\ ,
\end{array}
\]
with $\omega_8\equiv\frac{1}{\sqrt{6}}(u\bar u+d\bar d-2 s\bar s)$ and
$\omega_0\equiv\frac{1}{\sqrt{3}}(u\bar u+d\bar d+s\bar s)$.
Experimentally, the value of the vector mixing angle is measured to be above its ideal value
and is $\theta_V=(38.7\pm 0.2)^\circ$ \protect\cite{Dolinsky:vq}.
The value obtained from a quadratic Gell-Mann--Okubo mass formula for vectors mesons is
$\theta_V=(38.6\pm 0.4)^\circ$ \protect\cite{Eidelman:2004wy}.}
\begin{equation}
\label{FVPgamma00}
\begin{array}{l}
F_{\rho\eta\gamma}(0,0)=\frac{\sqrt{3}}{4\pi^2}
\frac{f^0_{\eta^\prime}-\sqrt{2}f^8_{\eta^\prime}}
{f^0_{\eta^\prime}f^8_\eta-f^8_{\eta^\prime}f^0_\eta}\ ,\\[1ex]
F_{\rho\eta^\prime\gamma}(0,0)=\frac{\sqrt{3}}{4\pi^2}
\frac{f^0_\eta-\sqrt{2}f^8_\eta}
{f^0_\eta f^8_{\eta^\prime}-f^8_\eta f^0_{\eta^\prime}}\ ,\\[1ex]
F_{\omega\eta\gamma}(0,0)=\frac{1}{2\sqrt{2}\pi^2}
\frac{(c\theta_V-s\theta_V/\sqrt{2})f^0_{\eta^\prime}-
      s\theta_V f^8_{\eta^\prime}}
{f^0_{\eta^\prime}f^8_\eta-f^8_{\eta^\prime}f^0_\eta}\ ,\\[1ex]
F_{\omega\eta^\prime\gamma}(0,0)=\frac{1}{2\sqrt{2}\pi^2}
\frac{(c\theta_V-s\theta_V/\sqrt{2})f^0_\eta-s\theta_V f^8_\eta}
{f^0_\eta f^8_{\eta^\prime}-f^8_\eta f^0_{\eta^\prime}}\ ,\\[1ex]
F_{\phi\eta\gamma}(0,0)=-\frac{1}{2\sqrt{2}\pi^2}
\frac{(s\theta_V+c\theta_V/\sqrt{2})f^0_{\eta^\prime}+
      c\theta_V f^8_{\eta^\prime}}
{f^0_{\eta^\prime}f^8_\eta-f^8_{\eta^\prime}f^0_\eta}\ ,\\[1ex]
F_{\phi\eta^\prime\gamma}(0,0)=-\frac{1}{2\sqrt{2}\pi^2}
\frac{(s\theta_V+c\theta_V/\sqrt{2})f^0_\eta+c\theta_V f^8_\eta}
{f^0_\eta f^8_{\eta^\prime}-f^8_\eta f^0_{\eta^\prime}}\ .\\[1ex]
\end{array}
\end{equation}
Using their analytic properties, we can express these form factors by
dispersion relations in the momentum of the vector current, which are then
saturated with the lowest-lying resonances:
\begin{equation}
\label{VMD}
F_{VP\gamma}(0,0)=\frac{f_V}{m_V}g_{VP\gamma}+\cdots\ ,
\end{equation}
where the dots stand for higher resonances and multiparticle contributions
to the correlation function.
In the following we assume vector meson dominance (VMD) and thus neglect these
contributions (see Ref.~\cite{Ball:1995zv} for further details).

The $f_V$ are the  leptonic decay constants of the vector mesons, and defined by
\begin{equation}
\label{lepdeccon}
\langle0|J_\mu^V|V(p,\lambda)\rangle=m_V f_V \varepsilon_\mu^{(\lambda)}(p)\ ,
\end{equation}
where $m_V$ and $\lambda$ are the mass and the helicity state of the vector meson.
The $f_V$ can be determined from the experimental decay rates \cite{Eidelman:2004wy} via
\begin{equation}
\label{GVee}
\Gamma(V\rightarrow e^+e^-)=\frac{4\pi}{3}\alpha^2\frac{f_V^2}{m_V}c_V^2\ ,
\end{equation}
with $c_V
=(\frac{1}{\sqrt{2}},\frac{s\theta_V}{\sqrt{6}},\frac{c\theta_V}{\sqrt{6}})$
for $V=\rho, \omega, \phi$.
The experimental values are
\begin{equation}
\label{fVexp}
\begin{array}{c}
f_{\rho^0}=(221\pm 2)\ \mbox{MeV}\ ,\\[1ex]
f_{\omega}=(180\pm 3)\ \mbox{MeV}\ ,\\[1ex]
f_{\phi}=(239\pm 4)\ \mbox{MeV}\ .
\end{array}
\end{equation}

Finally, we introduce the vertex couplings $g_{VP\gamma}$, which are just the
on-shell $V$-$P$ electromagnetic form factors:
\begin{equation}
\label{gVPgamma}
\langle P(p_P)|J_\mu^{\rm EM}|V(p_V,\lambda)\rangle|_{(p_V-p_P)^2=0}=
-g_{VP\gamma}\epsilon_{\mu\nu\alpha\beta}
p_P^\nu p_V^\alpha\varepsilon_V^\beta(\lambda)\ .
\end{equation}
The decay widths of $P\rightarrow V\gamma$ and $V\rightarrow P\gamma$ are
\begin{equation}
\label{GVPgamma}
\begin{array}{l}
\Gamma(P\rightarrow V\gamma)=
\frac{\alpha}{8}g_{VP\gamma}^2\left(\frac{m_P^2-m_V^2}{m_P}\right)^3\ ,\\[1ex]
\Gamma(V\rightarrow P\gamma)=
\frac{\alpha}{24}g_{VP\gamma}^2\left(\frac{m_V^2-m_P^2}{m_V}\right)^3\ .
\end{array}
\end{equation}

\TABLE{
\centering
{\scriptsize
\begin{tabular}{lllll}
\hline\hline\\
$V$ & $P$ & $g_{VP\gamma}$ (th.) & & $g_{VP\gamma}$ (exp.) \\[2ex]
\hline\\
$\rho$ & $\eta$ & 
$\frac{\sqrt{3} m_\rho}{4\pi^2 f_\rho}\frac{c\theta_0/f_8-\sqrt{2}s\theta_8/f_0}
{c\theta_0c\theta_8+s\theta_8s\theta_0}$
& $=(1.48\pm 0.08)$ GeV$^{-1}$ & $(1.59\pm 0.11)$ GeV$^{-1}$ \\[1ex]
$\rho$ & $\eta^\prime$ &
$\frac{\sqrt{3} m_\rho}{4\pi^2 f_\rho}\frac{s\theta_0/f_8+\sqrt{2}c\theta_8/f_0}
{c\theta_0c\theta_8+s\theta_8s\theta_0}$
& $=(1.23\pm 0.08)$ GeV$^{-1}$ & $(1.35\pm 0.06)$ GeV$^{-1}$ \\[1ex]
$\omega$ & $\eta$ &
$\frac{m_\omega}{2\sqrt{2}\pi^2 f_\omega}
\frac{(c\theta_V-s\theta_V/\sqrt{2})c\theta_0/f_8-s\theta_V s\theta_8/f_0}
{c\theta_0c\theta_8+s\theta_8s\theta_0}$
& $=(0.57\pm 0.04)$ GeV$^{-1}$ & $(0.46\pm 0.02)$ GeV$^{-1}$ \\[1ex]
$\omega$ & $\eta^\prime$ & 
$\frac{m_\omega}{2\sqrt{2}\pi^2 f_\omega}
\frac{(c\theta_V-s\theta_V/\sqrt{2})s\theta_0/f_8+s\theta_V c\theta_8/f_0}
{c\theta_0c\theta_8+s\theta_8s\theta_0}$
& $=(0.56\pm 0.04)$ GeV$^{-1}$ & $(0.46\pm 0.03)$ GeV$^{-1}$ \\[1ex]
$\phi$ & $\eta$ & 
$-\frac{m_\phi}{2\sqrt{2}\pi^2 f_\phi}
\frac{(s\theta_V+c\theta_V/\sqrt{2})c\theta_0/f_8+c\theta_V s\theta_8/f_0}
{c\theta_0c\theta_8+s\theta_8s\theta_0}$ 
& $=(-0.76\pm 0.04)$ GeV$^{-1}$ & $(-0.690\pm 0.008)$ GeV$^{-1}$ \\[1ex]
$\phi$ & $\eta^\prime$ & 
$-\frac{m_\phi}{2\sqrt{2}\pi^2 f_\phi}
\frac{(s\theta_V+c\theta_V/\sqrt{2})s\theta_0/f_8-c\theta_V c\theta_8/f_0}
{c\theta_0c\theta_8+s\theta_8s\theta_0}$
& $=(0.86\pm 0.05)$ GeV$^{-1}$ & $(0.71\pm 0.04)$ GeV$^{-1}$ \\[2ex]
\hline\hline
\end{tabular}
}
\caption{Theoretical and experimental values of the on-shell $V$-$(\eta,\eta^\prime)$
electromagnetic vertex couplings in the octet-singlet $\eta$-$\eta^\prime$
mixing angle scheme. 
For $g_{VP\gamma}$ (th.) we give the experimental errors coming from the decay constants
$f_{P,V}$ and the mixing angle values $\theta_8$ and $\theta_0$.
We use $\theta_V=(38.7\pm 0.2)^\circ$ for the $\phi$-$\omega$ mixing angle.
Experimental values are taken from {\protect\cite{Eidelman:2004wy}}.}
\label{table1}
}

Eq.~(\ref{VMD}) allows us to identify the $g_{VP\gamma}$ couplings defined in
(\ref{gVPgamma}) with the form factors $F_{VP\gamma}(0,0)$ listed in (\ref{FVPgamma00}).
The couplings are expressed in terms of the octet and singlet mixing angles $\theta_8$ and $\theta_0$,
the pseudoscalar decay constants $f_8$ and $f_0$, the $\phi$-$\omega$ mixing angle $\theta_V$,
and the corresponding vector decay constants $f_V$.
These theoretical expressions are shown in Table \ref{table1}.
We also include a numerical prediction for each coupling that should be compared with the 
experimental values extracted from (\ref{GVPgamma}) and Ref.~\cite{Eidelman:2004wy}.
In the numerical analysis we have taken into account a value for the vector mixing angle of
$\theta_V=(38.7\pm 0.2)^\circ$ \cite{Dolinsky:vq}. 
Our predictions are obtained from the results in Eq.~(\ref{fitresult134}).
The error quoted in Table \ref{table1} does not reflect the full theoretical uncertainty,
but namely propagates the errors from (\ref{fitresult134}) and (\ref{fVexp}).
The agreement between our theoretical predictions and the experimental values is quite
remarkable with exceptions in the $\omega\eta\gamma$ and $\omega\eta^\prime\gamma$ cases.
However, these two couplings merit some explanation.
On one case, the experimental value for $g_{\omega\eta\gamma}$ has changed from 
$(0.53\pm 0.05)$ GeV$^{-1}$ of the PDG'02 \cite{Hagiwara:2002fs} to the current
$(0.46\pm 0.02)$ GeV$^{-1}$ due to the exclusion of the measurement based on 
$e^+e^-\to\eta\gamma$ by Dolinsky \textit{et.~al.}~\cite{Dolinsky:vq}.
On the other case, the $g_{\omega\eta^\prime\gamma}$ coupling is rather sensitive to the
$\phi$-$\omega$ mixing angle;
for instance setting $\theta_V$ to the ideal mixing value of $35.3^\circ$
reduces the coupling by a 10\%.
As seen from Table \ref{table1}, the predictions for the decays $\phi\to (\eta,\eta^\prime)\gamma$,
which are the best measured and are largely independent of $\theta_V$,
are in good agreement with data.
It is worth noting that the comparison shown in Table \ref{table1} is performed with $f_8=1.34 f_\pi$.
Using instead $f_8=1.28 f_\pi$ one obtains similar numbers with the exception of 
$g_{\phi\eta\gamma}=(-0.80\pm 0.04)$ GeV$^{-1}$ and
$g_{\phi\eta^\prime\gamma}=(0.91\pm 0.06)$ GeV$^{-1}$ which are worse fitted.
In this sense, the experimental data seem to prefer a value for $f_8$ higher than the one
predicted by standard $\chi$PT.
Once again, the 2-angle calculation is the relevant one.
For completeness however, we have included in the table below same comparison but fixing 
$\theta_8=\theta_0\equiv\theta$.
In this case, the results are very similar for $f_8=1.28 f_\pi$ and $f_8=1.34 f_\pi$
but in both cases the $\chi^2/\textrm{d.o.f.}$~is increased by a factor of 3 as compared to the value
in the two mixing angle scheme.
In particular, $g_{\phi\eta^\prime\gamma}$ are fitted to
$(1.21\pm 0.07)$ GeV$^{-1}$ and $(1.20\pm 0.06)$ GeV$^{-1}$ respectively,
which are in clear contradiction with data.

\TABLE{
\centering
{\scriptsize
\begin{tabular}{ccc|ccc}
\hline\hline
&&&&&\\
Assumptions & Results & $\chi^2/\textrm{d.o.f.}$ &
Assumptions & Results & $\chi^2/\textrm{d.o.f.}$\\[1ex]\hline
&&&&&\\
$\theta_8$ and $\theta_0$ free  & $\theta_8=(-22.5\pm 1.3)^\circ$   & 42.3/6&
$\theta_8=\theta_0\equiv\theta$ & $\theta=(-16.9\pm 1.2)^\circ$     & 81.4/7\\[1ex]
$f_8=1.28 f_\pi$            & $\theta_0=(-8.0\pm 1.4)^\circ$    &   &
$f_8=1.28 f_\pi$            & $f_0=(1.15\pm 0.03) f_\pi$        &   \\[1ex]
$\theta_V=(38.7\pm 0.2)^\circ$  & $f_0=(1.21\pm 0.03) f_\pi$        &       &
$\theta_V=(38.7\pm 0.2)^\circ$  &                   &   \\[1ex]\hline
&&&&&\\
$\theta_8$ and $\theta_0$ free  & $\theta_8=(-22.9\pm 1.3)^\circ$   & 31.2/6&
$\theta_8=\theta_0\equiv\theta$ & $\theta=(-16.7\pm 1.1)^\circ$     & 77.8/7\\[1ex]
$f_8=1.34 f_\pi$            & $\theta_0=(-6.6\pm 1.4)^\circ$    &       &
$f_8=1.34 f_\pi$            & $f_0=(1.16\pm 0.03) f_\pi$        &   \\[1ex]
$\theta_V=(38.7\pm 0.2)^\circ$  & $f_0=(1.23\pm 0.03) f_\pi$        &       &
$\theta_V=(38.7\pm 0.2)^\circ$  &                   &   \\[1ex]\hline
&&&&&\\
$\theta_8$ and $\theta_0$ free  & $\theta_8=(-23.8\pm 1.4)^\circ$   & 18.8/5&
$\theta_8=\theta_0\equiv\theta$ & $\theta=(-16.5\pm 1.2)^\circ$     & 77.4/6\\[1ex]
$f_8$ free              & $\theta_0=(-2.4\pm 1.9)^\circ$    &       &
$f_8$ free              & $f_8=(1.37\pm 0.05) f_\pi$        &   \\[1ex]
$\theta_V=(38.7\pm 0.2)^\circ$  & $f_8=(1.51\pm 0.05) f_\pi$        &       &
$\theta_V=(38.7\pm 0.2)^\circ$  & $f_0=(1.17\pm 0.04) f_\pi$        &   \\[1ex]
                & $f_0=(1.29\pm 0.04) f_\pi$        &       &
                &                   &   \\[1ex]\hline
&&&&&\\
$\theta_8$ and $\theta_0$ free  & $\theta_8=(-24.0\pm 1.6)^\circ$   & 18.6/4&
$\theta_8=\theta_0\equiv\theta$ & $\theta=(-15.7\pm 1.4)^\circ$     & 76.1/5\\[1ex]
$f_8$ free              & $\theta_0=(-2.5\pm 1.9)^\circ$    &       &
$f_8$ free              & $f_8=(1.37\pm 0.05) f_\pi$        &   \\[1ex]
$\theta_V$ free         & $f_8=(1.51\pm 0.05) f_\pi$        &       &
$\theta_V$ free         & $f_0=(1.17\pm 0.03) f_\pi$        &   \\[1ex]
                & $f_0=(1.29\pm 0.04) f_\pi$        &       &
                & $\theta_V=(36.5\pm 1.8)^\circ$    &       \\[1ex]
                & $\theta_V=(39.4\pm 2.2)^\circ$    &       &
                &                   &   \\[2ex]
\hline\hline
\end{tabular}
}
\caption{Results for the $\eta$-$\eta^\prime$ mixing angles and decay constants
in the octet-singlet basis of the two mixing angle scheme \textit{(left)}
and in the one mixing angle scheme \textit{(right)}.
For every fit, the theoretical assumptions taken, the set of numerical results,
and the value of the $\chi^2/\textrm{d.o.f.}$~are shown in the first, second and third column
respectively.
The fitted experimental data includes the decay widths of
$(\eta,\eta^\prime)\rightarrow\gamma\gamma$, $V\rightarrow P\gamma$, $P\rightarrow V\gamma$,
and the ratio $R_{J/\psi}$.}
\label{table2}
}

Table \ref{table1} constitutes one of the main results of our work.
Our analysis shows that the assumption of saturating the form factors
$F_{VP\gamma}$ by lowest-lying resonances is satisfactory (a conclusion 
already reached in Ref.~\cite{Ball:1995zv}),
and that the $\eta$-$\eta^\prime$ system described in the two mixing angle scheme
(octet-singlet basis) fits the data much better than the one mixing angle scheme does.
To quantify this improvement, we have performed various fits to the full set of
experimental data assuming, or not, the two mixing angle scheme of the $\eta$-$\eta^\prime$ system. 
The results are presented in Table \ref{table2}.
To check the consistency of our approach we have extended the fit in 
Eq.~(\ref{fitresult128}) or Eq.~(\ref{fitresult134})
to include all experimental data that account not only for the decay widths 
$(\eta,\eta^\prime)\to\gamma\gamma$ and the ratio $R_{J/\psi}$
but also the radiative decay widths of $V\rightarrow P\gamma$ and $P\rightarrow V\gamma$. 
The theoretical constraint $f_8=1.28 f_\pi$ or $1.34 f_\pi$ is relaxed in order to test
the dependence of the result on the value of this parameter.
The experimental constrain $\theta_V=(38.7\pm 0.2)^\circ$ is also relaxed to test the stability
of the fit.

As seen from Table \ref{table2}, a significant improvement in the $\chi^2/\textrm{d.o.f.}$~is
achieved when the constrain $\theta_8=\theta_0\equiv\theta$ is relaxed
(in the most favorable case the $\chi^2/\textrm{d.o.f.}$~is reduced by more than a factor of 3),
allowing us to show explicitly the improvement of our analysis using the two mixing angle scheme
with respect to the one using the one mixing angle scheme.
In general, the fits with $1.34 f_\pi$ are slightly better than those with $f_8=1.28 f_\pi$.
To leave $f_8$ as a free parameter does not make any substantial difference in the
one angle case while in the case of two mixing angles the value increases up to 
$f_8=(1.51\pm 0.05) f_\pi$, in disagreement with the expectations of $\chi$PT and
Large $N_c$ $\chi$PT.
However, the analysis of the parameter correlation coefficients reveals a strong positive
correlation between $f_8$ and $\theta_0$ ($+0.674$)
---the latter quantity being taken as a negative number in the present convention;
in other terms, the correlation to the \emph{absolute} value of $\theta_0$ is negative.
The correlations between $f_8$ and $\theta_8$ or $\theta_8$ and $\theta_0$ are negative and
much weaker ($-0.126$ and $-0.124$, respectively).
It is also observed that when $f_8$ is fixed, the remaining correlation coefficients are smaller.
This increase is translated into a considerably better fit.
Finally, it is seen in Table \ref{table2} that relaxing the experimental constrain
$\theta_V=(38.7\pm 0.2)^\circ$ does not produce any effect on the fits.

\TABLE{
\centering
{\scriptsize
\begin{tabular}{ccc|ccc}
\hline\hline
&&&&&\\
Assumptions & Results & $\chi^2/\textrm{d.o.f.}$ &
Assumptions & Results & $\chi^2/\textrm{d.o.f.}$\\[1ex]\hline
&&&&&\\
$\phi_q$ and $\phi_s$ free  & $\phi_q=(40.4\pm 1.2)^\circ$      & 34.6/6&
$\phi_q=\phi_s\equiv\phi$   & $\phi=(40.8\pm 0.9)^\circ$        & 34.9/7\\[1ex]
$f_q=f_\pi$             & $\phi_s=(41.3\pm 1.3)^\circ$      &   &
$f_q=f_\pi$             & $f_s=(1.66\pm 0.06) f_\pi$        &   \\[1ex]
$\phi_V=(3.4\pm 0.2)^\circ$ & $f_s=(1.66\pm 0.06) f_\pi$        &       &
$\phi_V=(3.4\pm 0.2)^\circ$ &                   &   \\[1ex]\hline
&&&&&\\
$\phi_q$ and $\phi_s$ free  & $\phi_q=(39.9\pm 1.3)^\circ$      & 18.8/5&
$\phi_q=\phi_s\equiv\phi$   & $\phi=(40.6\pm 0.9)^\circ$        & 19.4/6\\[1ex]
$f_q$ free              & $\phi_s=(41.4\pm 1.4)^\circ$      &       &
$f_q$ free              & $f_q=(1.10\pm 0.03) f_\pi$        &   \\[1ex]
$\phi_V=(3.4\pm 0.2)^\circ$ & $f_q=(1.09\pm 0.03) f_\pi$        &       &
$\phi_V=(3.4\pm 0.2)^\circ$ & $f_s=(1.66\pm 0.06) f_\pi$        &   \\[1ex]
                & $f_s=(1.66\pm 0.06) f_\pi$        &       &
                &                   &   \\[1ex]\hline
&&&&&\\
$\phi_q$ and $\phi_s$ free  & $\phi_q=(39.8\pm 1.3)^\circ$      & 18.6/4&
$\phi_q=\phi_s\equiv\phi$   & $\phi=(40.4\pm 1.0)^\circ$        & 19.3/5\\[1ex]
$f_q$ free              & $\phi_s=(41.2\pm 1.5)^\circ$      &       &
$f_q$ free              & $f_q=(1.10\pm 0.03) f_\pi$        &   \\[1ex]
$\phi_V$ free           & $f_q=(1.09\pm 0.03) f_\pi$        &       &
$\phi_V$ free           & $f_s=(1.66\pm 0.07) f_\pi$        &   \\[1ex]
                & $f_s=(1.67\pm 0.07) f_\pi$        &       &
                & $\phi_V=(4.1\pm 2.2)^\circ$       &       \\[1ex]
                & $\phi_V=(4.2\pm 2.1)^\circ$       &       &
                &                   &   \\[2ex]
\hline\hline
\end{tabular}
}
\caption{Results for the $\eta$-$\eta^\prime$ mixing angles and decay constants
in the quark-flavour basis of the two mixing angle scheme \textit{(left)}
and in the one mixing angle scheme \textit{(right)}.
The conventions are the same as in Table \protect\ref{table2}.}
\label{table3}
}

Up to now, we have shown in the octet-singlet basis the need for a two mixing angle scheme
in order to describe experimental data in a better way.
In the following, we proceed to perform the same kind of analysis but in the quark-flavour basis.
We will see that at the \emph{current experimental} accuracy, the two angles have compatible values in this case.
There is however no strong reason to impose this equality as a constraint.
See also above for a discussion of the respective effects of the strange quark mass and anomalies in the mixing.

In Table \ref{table3} we present the results of various fits taking into account
all experimental data available
---namely $(\eta, \eta^\prime)\to\gamma\gamma$, $V\to P\gamma$, $P\to V\gamma$,
and the ratio $R_{J/\psi}$---
and the corresponding theoretical expressions found in
Eqs.~(\ref{theowidthsquarkbasis})--(\ref{gVPgammaquarkbasis}) of the Appendix.
The theoretical constrain $f_q=f_\pi$ is relaxed to test the dependence of the result
on the value of this parameter and the same happens with the experimental constrain
$\phi_V=(3.4\pm 0.2)^\circ$.
As seen from Table \ref{table3}, there is no significant difference at the present experimental accuracy between the
$\chi^2/\textrm{d.o.f.}$ of the fits when data are described in terms of 
two mixing angles (in the quark-flavour basis) or if 
$\phi_q=\phi_s\equiv\phi$ is imposed.
It is however important to notice that the fit considerably improves when the parameter $f_q$ is left free.
In this case the value obtained $f_q=(1.10\pm 0.03)f_\pi$ is incompatible with the value in the 
large $N_c$ limit $f_q=f_\pi$.

\section{Discussion about the mixing parameters}
\label{discussion}
In this Section, we compare our best results for the pseudoscalar decay constants and mixing angles
in the octet-singlet and quark-flavour basis with the theoretical expectations of Large $N_c$ $\chi$PT
and previous phenomenological analyses.
The value of the mixing parameters extracted from our best fit in the two mixing angle scheme are
\begin{equation}
\label{finalresultsos}
\begin{array}{ll}
f_8=(1.51\pm 0.05)f_\pi\ ,\quad & \theta_8=(-23.8\pm 1.4)^\circ\ ,\\[1ex]
f_0=(1.29\pm 0.04)f_\pi\ ,\quad & \theta_0=(-2.4\pm 1.9)^\circ\ ,
\end{array}
\end{equation}
in the octet-singlet basis, and
\begin{equation}
\label{finalresultsqf}
\begin{array}{ll}
f_q=(1.09\pm 0.03)f_\pi\ ,\quad & \phi_q=(39.9\pm 1.3)^\circ\ ,\\[1ex]
f_s=(1.66\pm 0.06)f_\pi\ ,\quad & \phi_s=(41.4\pm 1.4)^\circ\ ,
\end{array}
\end{equation}
in the quark-flavour basis.
These values are extracted from a comparison with experimental data only assuming that
the pseudoscalar decay constants involved in the corresponding processes follow the
parametrization given in Eqs.~(\ref{defdecaybasisos}) and (\ref{defdecaybasisqf}), respectively.
At the present accuracy, our results satisfy the approximate relations existing between
the two different sets of mixing parameters\footnote{
The relations (\protect\ref{mixparosqf}) are obtained once
the non-strange and strange axial-vector currents are written in terms of
the octet and singlet ones (see Eq.~(\protect\ref{divaxialqfos}))
and are valid for $\phi_q=\phi_s\equiv\phi$.}
\cite{Feldmann:1998vh}:
\begin{equation}
\label{mixparosqf}
\begin{array}{ll}
f_8=\sqrt{1/3 f_q^2+2/3 f_s^2}\ ,\quad & \theta_8=\phi-\arctan(\sqrt{2}f_s/f_q)\ ,\\[2ex]
f_0=\sqrt{2/3 f_q^2+1/3 f_s^2}\ ,\quad & \theta_0=\phi-\arctan(\sqrt{2}f_q/f_s)\ .
\end{array}
\end{equation}

Large $N_c$ $\chi$PT predicts \cite{Leutwyler:1997yr,Kaiser:1998ds}:
\begin{equation}
\label{mixparthos}
\begin{array}{l}
f_8^2=\frac{4f_K^2-f_\pi^2}{3}\ ,\quad
f_0^2=\frac{2f_K^2+f_\pi^2}{3}+f_\pi^2\Lambda_1\ ,\\[1ex]
f_8 f_0\sin(\theta_8-\theta_0)=-\frac{2\sqrt{2}}{3}(f_K^2-f_\pi^2)\ ,
\end{array}
\end{equation}
in the octet-singlet basis and
\begin{equation}
\label{mixparthqf}
\begin{array}{l}
f_q^2=f_\pi^2+\frac{2}{3}f_\pi^2\Lambda_1\ ,\quad
f_s^2=2f_K^2-f_\pi^2+\frac{1}{3}f_\pi^2\Lambda_1\ ,\\[1ex]
f_q f_s\sin(\phi_q-\phi_s)=\frac{\sqrt{2}}{3}f_\pi^2\Lambda_1\ ,
\end{array}
\end{equation}
in the quark-flavour basis \cite{Feldmann:1999uf}.
These expressions are valid at next-to-leading order in the Large $N_c$ $\chi$PT
expansion where the octet-singlet (and quark-flavour) pseudoscalar decay constants
can be written in terms of the known $f_\pi$ and $f_K$ decay constants
and the unknown OZI-rule violating parameter $\Lambda_1$.
Using the experimental constrain $f_K=1.22 f_\pi$, one obtains\footnote{
A value of $f_8=1.34 f_\pi$ is obtained if chiral logs and higher order contributions
are also taken into account \protect\cite{Kaiser:1998ds}.}
\cite{Leutwyler:1997yr}:
$f_8=1.28 f_\pi$, $\theta_8=-20.5^\circ$, $f_0\simeq 1.25$, and $\theta_0\simeq -4^\circ$.
Our results in Eq.~(\ref{finalresultsos}) are quite in agreement with the former values
except for the case of $f_8$.
Note however that in Ref.~\cite{Leutwyler:1997yr} the value of $f_8$ is fixed from theory
while in our analysis it is fitted from a direct comparison with experimental data
where a positive correlation between $f_8$ and $\theta_0$ appears.
For $f_8=1.34 f_\pi$ the results of the fit are in perfect agreement with the predictions
from Large $N_c$ $\chi$PT even though the quality of the fit is slightly reduced.

As seen from Table \ref{table2},
if the constrain $f_8=1.28 f_\pi$ is imposed one gets a worse fit.
The same kind of comparison can be performed in the quark-flavour basis.
From the phenomenological analysis \cite{Feldmann:1998vh}
the values $f_q=(1.07\pm 0.02)f_\pi$, $f_s=(1.34\pm 0.06)f_\pi$ and $\phi=(39.3\pm 1.0)^\circ$
are obtained.
In this case, our results in Eq.~(\ref{finalresultsqf}) fairly agree with the exception of $f_s$
which clearly disagrees.
This difference may be due to the fact that the analysis in Ref.~\cite{Feldmann:1998vh}
is based on a different set of experimental data not including for instance the very precise
and recent $\phi\to (\eta, \eta^\prime)\gamma$ decays which are very dependent on $f_s$
(see Eq.~(\ref{gVPgammaquarkbasis})).

Our best results for the mixing parameters in Eqs.~(\ref{finalresultsos}) and (\ref{finalresultsqf})
can be used to check the consistency of Eqs.~(\ref{mixparthos}) and (\ref{mixparthqf}),
and therefore to test the reliability on the Large $N_c$ $\chi$PT framework.
Accordingly, our fitted values for $f_8$ and $f_0$ together with the third equation in (\ref{mixparthos})
can be used to get $\theta_8-\theta_0=(-13.7\pm 0.6)^\circ$ as a prediction for the difference
of the two mixing angles in the octet-singlet basis.
If one compares this prediction with our result $\theta_8-\theta_0=(-21.4\pm 2.4)^\circ$
a disagreement is again obtained.
Using $f_8=1.34 f_\pi$ and our fitted value for $f_0$ one gets $\theta_8-\theta_0=(-16.2\pm 0.4)^\circ$
to compare with our result $\theta_8-\theta_0=(-16.3\pm 1.9)^\circ$ (see Table \ref{table2}).
The second equation in (\ref{mixparthos}) may be used to get a prediction for the 
OZI-rule violating parameter $\Lambda_1$ once $f_0$ is provided.
Using our value for $f_0$ one gets $\Lambda_1=0.34\pm 0.10$.
$\Lambda_1$ can also be extracted from the set of equations (\ref{mixparthqf}).
From the fitted values for $f_{q,s}$ and $\phi_{q,s}$ the results
$\Lambda_1=0.32\pm 0.10$, $2.34\pm 0.60$ and $-0.10\pm 0.13$
are obtained using the first, second and third equation in (\ref{mixparthqf}), respectively.
Indeed, the same behaviour is also observed in the phenomenological analysis of
Ref.~\cite{Feldmann:1998vh}.
The values obtained for the mixing parameters in this analysis lead to values for $\Lambda_1$
that are incompatible with the constrain $\Lambda_1\equiv 0$ imposed in their fit.

The numbers obtained in the former discussion hint at a disagreement with Large $N_c$ $\chi$PT.
Some care should be taken however to qualify this statement.
On the one hand, the statistical significance is still limited,
but will obviously improve as critical channels are measured more acurately.
On the other hand, it is plainly clear that the critical information comes here from
radiative decays of vector mesons, where we use vector meson dominance.
Questions have been raised as to the consistency of vector resonances and VMD with
Chiral Perturbation Theory and short distance QCD cf.~e.g.~\cite{Knecht:2001xc}.
This latter remark should however be moderated by the fact that our approach was carefully tested
in the charged mesons sector (away from the eta mixing problems) \cite{Ball:1995zv}.

\section{Conclusions}
\label{conclusions}
In this work we have performed a phenomenological analysis on various decay processes
using a two mixing angle scheme for the $\eta$-$\eta^\prime$ system.
First we have used the radiative decays $(\eta,\eta^\prime)\to\gamma\gamma$ together with
the ratio $R_{J/\psi}$ to fit the values of the pseudoscalar decay constants and the mixing angles
in both the octet-singlet and the quark-flavour basis.
Using the description of vector meson decays in terms of their relation with the
$AVV$ triangle anomaly, a theoretical prediction for the $g_{VP\gamma}$ couplings have been derived.
The agreement between our theoretical predictions and the experimental values is very good
and can be considered as a consistency check of the whole approach.
Second we have extend our analysis to include the $V\to P\gamma$ and $P\to V\gamma$ decays
in the fits showing that a two mixing angle description in the octet-singlet basis is 
definitely required in order to achieve good agreement with experimental data.
On the contrary, in the quark-flavour basis and with the present experimental accuracy
a one mixing angle description of the processes is still enough to reach agreement.
Finally, we have compared our best fitted mixing parameters with the predictions from
Large $N_c$ $\chi$PT showing a possible discrepancy with this framework.
Higher accuracy data and more refined theoretical analyses would contribute to clarify the
preceding issue.

\section*{Acknowledgements}
The authors acknowledge valuable discussions with J.-M.~Gerard and E.~Kou.
R.E.~also acknowledges F.~S\'anchez for useful discussions.
This work is  partly supported by the Ramon y Cajal program (R.E.),
the Ministerio de Ciencia y Tecnolog\'{\i}a and FEDER, FPA2002-00748EU,
and the EU, HPRN-CT-2002-00311, EURIDICE network,
as well as the IISN (Belgian French community) and the 
Inter-University Attraction Pole V/27 (Belgian federal science policy office). 

\appendix
\section{Theoretical expressions in the quark-flavour basis}
In the quark-flavour basis,
the interpolating fields $\eta$ and $\eta^\prime$ are expressed as
\begin{equation}
\label{interfieldsquarkbasis}
\begin{array}{c}
\eta(x)=\frac{1}{m_\eta^2}
\frac{f^s_{\eta^\prime}\partial^\mu A_\mu^q(x)-
      f^q_{\eta^\prime}\partial^\mu A_\mu^s(x)}
{f^s_{\eta^\prime}f^q_\eta-f^q_{\eta^\prime}f^s_\eta}\ ,\qquad
\eta^\prime(x)=\frac{1}{m_{\eta^\prime}^2}
\frac{f^s_\eta\partial^\mu A_\mu^q(x)-
      f^q_\eta\partial^\mu A_\mu^s(x)}
{f^s_\eta f^q_{\eta^\prime}-f^q_\eta f^s_{\eta^\prime}}\ ,
\end{array}
\end{equation}
the $(\eta, \eta^\prime)\rightarrow\gamma\gamma$ decay widths as
\begin{equation}
\label{theowidthsquarkbasis}
\begin{array}{c}
\Gamma(\eta\rightarrow\gamma\gamma)=
\frac{\alpha^2 m_\eta^3}{32\pi^3}\left(
\frac{C_{q}f^s_{\eta^\prime}-C_{s}f^q_{\eta^\prime}}
{f^s_{\eta^\prime}f^q_\eta-f^q_{\eta^\prime}f^s_\eta}\right)^2=
\frac{\alpha^2 m_\eta^3}{32\pi^3}\left(
\frac{C_{q}c\phi_s/f_q-C_{s}s\phi_q/f_s}
{c\phi_s c\phi_q+s\phi_q s\phi_s}\right)^2\ ,\\[2ex]
\Gamma(\eta^\prime\rightarrow\gamma\gamma)=
\frac{\alpha^2 m_{\eta^\prime}^3}{32\pi^3}\left(
\frac{C_{q}f^s_\eta-C_{s}f^q_\eta}
{f^s_\eta f^q_{\eta^\prime}-f^q_\eta f^s_{\eta^\prime}}\right)^2=
\frac{\alpha^2 m_{\eta^\prime}^3}{32\pi^3}\left(
\frac{C_{q}s\phi_s/f_q+C_{s}c\phi_q/f_s}
{c\phi_s c\phi_q+s\phi_q s\phi_s}\right)^2\ ,
\end{array}
\end{equation}
with $C_{q}=5/3$ and $C_{s}=\sqrt{2}/3$,
and the ratio $R_{J/\psi}$ as
\begin{equation}
\label{RJpsiourtheoquarkbasis}
R_{J/\psi}=
\left|\frac{m_{\eta^\prime}^2 f^q_{\eta^\prime}}{m_\eta^2 f^q_\eta}\right|^2
\left(\frac{p_{\eta^\prime}}{p_\eta}\right)^3=
\tan^2\phi_{q}\left(\frac{m_{\eta^\prime}}{m_\eta}\right)^4
\left(\frac{p_{\eta^\prime}}{p_\eta}\right)^3\ .
\end{equation}

The form factors $F_{VP\gamma}(0,0)$ are written in terms of the 
pseudoscalar decay constants ($f_{q}, f_{s}$) and the $\phi$-$\omega$ mixing angle 
in the quark-flavour basis
($\phi_{V}=\theta_{V}-\arctan\frac{1}{\sqrt{2}}$):
\begin{equation}
\label{FVPgamma00quarkbasis}
\begin{array}{ll}
F_{\rho\eta\gamma}(0,0)=\frac{3}{4\pi^2}\frac{f^s_{\eta^\prime}}
{f^s_{\eta^\prime}f^q_\eta-f^q_{\eta^\prime}f^s_\eta}\ ,\quad &
F_{\rho\eta^\prime\gamma}(0,0)=\frac{3}{4\pi^2}\frac{f^s_\eta}
{f^s_\eta f^q_{\eta^\prime}-f^q_\eta f^s_{\eta^\prime}}\ ,\\[2ex]
F_{\omega\eta\gamma}(0,0)=\frac{1}{4\pi^2}
\frac{c\phi_V f^s_{\eta^\prime}-2s\phi_V f^q_{\eta^\prime}}
{f^s_{\eta^\prime}f^q_\eta-f^q_{\eta^\prime}f^s_\eta}\ ,\quad &
F_{\omega\eta^\prime\gamma}(0,0)=\frac{1}{4\pi^2}
\frac{c\phi_V f^s_\eta-2s\phi_V f^q_\eta}
{f^s_\eta f^q_{\eta^\prime}-f^q_\eta f^s_{\eta^\prime}}\ ,\\[2ex]
F_{\phi\eta\gamma}(0,0)=-\frac{1}{4\pi^2}
\frac{s\phi_V f^s_{\eta^\prime}+2c\phi_V f^q_{\eta^\prime}}
{f^s_{\eta^\prime}f^q_\eta-f^q_{\eta^\prime}f^s_\eta}\ ,\quad &
F_{\phi\eta^\prime\gamma}(0,0)=-\frac{1}{4\pi^2}
\frac{s\phi_V f^s_\eta+2c\phi_V f^q_\eta}
{f^s_\eta f^q_{\eta^\prime}-f^q_\eta f^s_{\eta^\prime}}\ .
\end{array}
\end{equation}
Finally, the vertex couplings $g_{VP\gamma}$ in this basis are
\begin{equation}
\label{gVPgammaquarkbasis}
\begin{array}{l}
g_{\rho\eta\gamma}=
\frac{3 m_\rho}{4\pi^2 f_\rho}\frac{c\phi_s/f_q}{c\phi_sc\phi_q+s\phi_qs\phi_s}\simeq
\frac{3 m_\rho}{4\pi^2 f_\rho}\frac{c\phi}{f_q}\ ,\\[2ex]
g_{\rho\eta^\prime\gamma}=
\frac{3 m_\rho}{4\pi^2 f_\rho}\frac{s\phi_s/f_q}{c\phi_sc\phi_q+s\phi_qs\phi_s}\simeq
\frac{3 m_\rho}{4\pi^2 f_\rho}\frac{s\phi}{f_q}\ ,\\[2ex]
g_{\omega\eta\gamma}=
\frac{m_\omega}{4\pi^2 f_\omega}
\frac{c\phi_V c\phi_s/f_q-2s\phi_V s\phi_q/f_s}{c\phi_sc\phi_q+s\phi_qs\phi_s}\simeq
\frac{m_\omega}{4\pi^2 f_\omega}\left(c\phi_V\frac{c\phi}{f_q}-2s\phi_V\frac{s\phi}{f_s}\right)\ ,\\[2ex]
g_{\omega\eta^\prime\gamma}= 
\frac{m_\omega}{4\pi^2 f_\omega}
\frac{c\phi_V s\phi_s/f_q+2s\phi_V c\phi_q/f_s}{c\phi_sc\phi_q+s\phi_qs\phi_s}\simeq
\frac{m_\omega}{4\pi^2 f_\omega}\left(c\phi_V\frac{s\phi}{f_q}+2s\phi_V\frac{c\phi}{f_s}\right)\ ,\\[2ex]
g_{\phi\eta\gamma}=
-\frac{m_\phi}{4\pi^2 f_\phi}
\frac{s\phi_V c\phi_s/f_q+2c\phi_V s\phi_q/f_s}{c\phi_sc\phi_q+s\phi_qs\phi_s}\simeq
-\frac{m_\phi}{4\pi^2 f_\phi}\left(s\phi_V\frac{c\phi}{f_q}+2c\phi_V\frac{s\phi}{f_s}\right)\ ,\\[2ex]
g_{\phi\eta^\prime\gamma}= 
-\frac{m_\phi}{4\pi^2 f_\phi}
\frac{s\phi_V s\phi_s/f_q-2c\phi_V c\phi_q/f_s}{c\phi_sc\phi_q+s\phi_qs\phi_s}\simeq
-\frac{m_\phi}{4\pi^2 f_\phi}\left(s\phi_V\frac{s\phi}{f_q}-2c\phi_V\frac{c\phi}{f_s}\right)\ ,
\end{array}
\end{equation}
where the approximations are valid for $\phi_q=\phi_s\equiv\phi$.

\end{document}